\documentclass[epj,referee]{svjour}

\usepackage{graphicx}
\usepackage{xcolor}
\usepackage{appendix}
\usepackage{hyperref}
\usepackage{booktabs}
\usepackage{adjustbox}

\newcommand{\sq}{\sqrt{s_{\mathrm{NN}}}}
\newcommand{\vslope}{dv_{1}/d\eta}
\newcommand{\pt}{p_\mathrm{T}}

\begin{document}

\title{System size dependence of charged hadrons directed flow at $\sqrt{s_{NN}}$ = 200 GeV using a multi-phase transport model}
\author{Kishora Nayak\inst{1}, Vipul Bairathi\inst{2}}
\institute{Department of Physics, Panchayat College, Sambalpur University, Bargarh-768028, Odisha, India 
\and Instituto de Alta Investigación, Universidad de Tarapacá, Casilla 7D Arica 1000000, Chile}

\offprints{Vipul Bairathi} 
\mail{vipul.bairathi@gmail.com}

\date{Received: date / Revised version: date} 

\abstract{
The directed flow ($v_1$) of charged hadrons ($h^{\pm}$) in symmetric collision systems (O+O, Cu+Cu, Zr+Zr, Ru+Ru, Au+Au, and U+U) at $\sqrt{s_{\mathrm{NN}}} =$ 200 GeV using string-melting version of A Multiphase Transport (AMPT-SM) model is reported. The $v_1$ as a function of pseudo-rapidity ($\eta$) is obtained for transverse momentum ($p_{\mathrm{T}}$) ranges of 0.2-2.0 GeV/$c$ and 2.0-5.0 GeV/$c$. The dependence of $v_1$-slope ($dv_1/d\eta$) at mid-rapidity on $p_{\mathrm{T}}$ range, collision centrality, and system size are discussed particularly in the context of the hard-soft asymmetry in the flow profiles of produced particles. In the AMPT-SM model, a system size independence of the magnitude of $dv_1/d\eta$ between Cu+Cu and Au+Au collisions at low-$\pt$ is observed, and this finding is similar to the observation from the STAR experiment at $\sqrt{s_{\mathrm{NN}}} =$ 200 GeV. In contrast, a strong centrality and system size dependence, with the opposite sign of $dv_1/d\eta$, is found for the high-$p_{\mathrm{T}}$ charged hadrons. The AMPT-SM model demonstrates a clear violation of the expected scaling of charged hadrons $(dv_1/d\eta)/A^{1/3}$ across different colliding systems.}

\PACS{
      {25.75.-q}{Relativistic heavy-ion collisions}   \and
      {25.75.Ld}{Collective flow, relativistic collisions}
     } 
\titlerunning{system size dependence of directed flow}
\authorrunning{K. Nayak {\it et al.}} 
\maketitle
\clearpage

\section{Introduction}
\label{sec:Intro}
Experimental findings from relativistic heavy-ion collisions provide compelling evidence for a deconfined quark-gluon plasma (QGP) matter~\cite{exqgp1,exqgp2,exqgp3,exqgp4,exqgp5,exqgp6,exqgp7}. High-energy heavy-ion collision programs aim to understand the properties of QGP and the initial conditions of these collisions. Azimuthal anisotropic flow is particularly sensitive to the early stages of collisions, where the collision dynamics is expected to be dominated by QGP matter~\cite{flow1}. The anisotropic flow is characterized by the coefficients $v_n$ in the Fourier expansion of the particle azimuthal angle distribution relative to the collision symmetry planes~\cite{flow2,flow3}. 

Directed flow $v_1$, is the first harmonic coefficient, consists of odd and even pseudo-rapidity components. The odd component, $v_{1}^{\mathrm{odd}}(\eta) = -v_{1}^{\mathrm{odd}}(-\eta)$, is associated with the collective sideward motion of produced particles relative to the collision symmetry plane. It is believed to be generated during the nuclear passage time, making it sensitive to the early stage collision dynamics~\cite{dflow1}. However, the even component, $v_{1}^{\mathrm{even}}(\eta) = v_{1}^{\mathrm{even}}(-\eta)$, is not correlated to the collision symmetry plane, and originates from the event-by-event fluctuations in the positions of participant nucleons and the number of spectators in the collision. This study focuses on $v_{1}^{\mathrm{odd}}(\eta)$, which will be implicitly referred to as $v_{1}$ and its slope at mid-pseudorapidity, $\vslope|_{\eta=0}$ as $\vslope$ throughout the paper.

\begin{figure}[!htbp]
\centering
\includegraphics[scale=0.5]{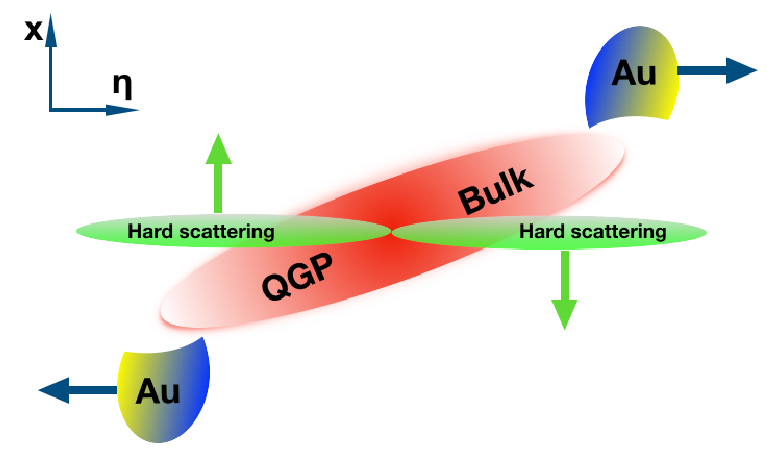}
\caption{(Color online) Schematic illustration of the orientation of the initial tilted QGP bulk relative to the symmetric hard scattering distribution profile of particle production projected onto the reaction plane (x, z)~\cite{dflow2,dflow3}.}
\label{fig:cartoon}
\end{figure}

Directed flow is sensitive to the release of compressional energy and the tilted shape of the matter created during heavy-ion collisions. The tilt results from asymmetries in the number of participant nucleons moving forward and backward in the transverse plane~\cite{flow4,flow5,flow6}. The bulk of QGP, which exhibits a tilt in rapidity, is primarily composed of low-$\pt$ hadrons produced through soft processes~\cite{dflow2,dflow3}. In contrast, the production profile of high-$\pt$ hadrons is symmetric in rapidity, as shown in Fig.~\ref{fig:cartoon}. The asymmetry between hadrons produced via hard and soft processes in the initial state results in a negative $v_{1}$ for hard partons. Therefore, studying the $\pt$-dependence of $v_{1}$ and $\vslope$ is necessary to understand the hard-soft asymmetry in the flow profile of particles produced in heavy-ion collisions.
The $v_1$ of high-$\pt$ hadrons is created in the early stages of the collisions and retain information of the particle production with the medium evolution until the freeze-out. Thus, it is possible to investigate the early time thermalization or pre-equilibrium stage using the $v_1$.

Extensive measurements of $v_{1}$ have been carried out at both the RHIC and LHC~\cite{exdflow1,exdflow2,exdflow3,exdflow4,exdflow5,exdflow6,exdflow7,exdflow8,exdflow9,exdflow10}. Model studies indicate that $v_1$ is sensitive to the shear viscosity of hot QCD matter. Furthermore, $v_1$ provides strong constraints on baryon stopping and can serve as a probe for the equation of state in heavy-ion collisions \cite{dflow4,dflow5,dflow6,dflow7,ampt_30AGeV}. The dependence of $v_1$ on system size, with varying collision geometries, is expected to provide critical tests of the parametric dependence of $v_1$ and insights into the mechanisms underlying charged particle production. 

In this paper, we report a comprehensive study of charged hadron $v_{1}$ at low- and high-$\pt$ in O+O, Cu+Cu, Zr+Zr, Ru+Ru, Au+Au, and U+U collisions at $\sq$ = 200 GeV using string melting version of the AMPT model to understand the effect of system size on $v_1$. The slope of $v_1$ is obtained at both low- and high-$\pt$ to gain insight into the hard-soft asymmetry in the particle production profile. Additionally, we examine the scaling properties of $\vslope$ across different colliding systems. 

The paper is structured into the following sections. Section~\ref{sec:model} provides a brief overview of the AMPT model, including modifications made to incorporate deformation in the structure of colliding nuclei. Section~\ref{subsec:analysis} presents the analysis and results, highlighting the dependence of system size on directed flow and its slope in heavy-ion collisions. The particle production profile is also discussed by studying $v_1$ at low- and high-$\pt$. Finally, section~\ref{sub:summary} summarizes the findings and provides an outlook for future research.

\section{The AMPT model}
\label{sec:model}
The AMPT model comprises four key components: initial conditions, partonic interactions, conversion to hadronic matter, and hadronic interactions~\cite{AMPT}. Initial conditions are derived from the HIJING model and include distributions of minijet partons and soft string excitations~\cite{HIJING}. The model employs Zhang’s parton cascade (ZPC) for parton scatterings, focusing on two-body interactions based on pQCD cross-sections~\cite{ZPC}. In the default version of AMPT, partons are combined back with parent strings and converted to hadrons using the Lund string fragmentation model. In the string melting version, a quark coalescence model is applied. The dynamics of the resulting hadronic matter are governed by a cascade based ART model~\cite{ART}. 

The nucleons inside the nuclei in AMPT is modeled using the Wood-Saxon (WS) distribution function defined as follows, \begin{equation}
\rho(r,\theta) = \frac{\rho_{0}}{1+e^{[\lbrace r - R(\theta,\phi)\rbrace/a]}},
\end{equation}
where $\rho_{0}$ is the normal nuclear density, $r$ is the distance from the center of the nucleus, $a$ is the surface diffuseness parameter, and $R(\theta,\phi)$ is the parameter characterizing the deformation of the nucleus, 
\begin{equation} 
R(\theta,\phi) = R_{0}[1 + \beta_{2}Y_{2,0}(\theta,\phi) + \beta_{3}Y_{3,0}(\theta,\phi)].
\end{equation} 
$R_{0}$ represents the radius parameter, $\beta_{2}$ and $\beta_{3}$ are the quadrupole and octuple deformities, and $Y_{l,m}(\theta,\phi)$ are the spherical harmonics.

\begin{table}[!htbp]
\begin{center}
\caption{Wood-Saxon parameters for various nuclei in the AMPT-SM model.}
\label{tab1}
\begin{adjustbox}{width=\columnwidth,center}
\begin{tabular}{ccccccc}
\hline
\hline
Parameter 	& O  	& Cu 	& Zr 	& Ru 	& Au 	& U		\\
\hline
$Z$         & 8  	& 29 	& 40 	& 44 	& 79  	& 92	\\
$A$         & 16 	& 63 	& 96 	& 96 	& 197 	& 238	\\
$R_{0}$     & 2.608 & 4.214 & 5.090 & 5.090 & 6.380 & 6.810 \\
$a$         & 0.513 & 0.586 & 0.520 & 0.460 & 0.535 & 0.550 \\
$\beta_{2}$ & 0 	& 0 	& 0.060 & 0.162	& 0 	& 0.280 \\
$\beta_{3}$ & 0 	& 0 	& 0.200	& 0		& 0 	& 0 	\\
\hline
\hline
\end{tabular}
\end{adjustbox}
\end{center}
\end{table}

We utilized an improved version of the coalescence AMPT-SM model, incorporating a partonic cross-section of 1.5 mb. Additionally, we effectively turned off final state hadronic interactions by setting the hadron cascade time, $t_{max} =$ 0.4 $fm/c$, to study the effect of system size on directed flow. We generated event for various collision systems, including O+O, Cu+Cu, Ru+Ru, Zr+Zr, Au+Au, and U+U at $\sqrt{s_{\mathrm {NN}}}$ = 200 GeV with the corresponding WS parameters, as shown in Table~\ref{tab1}. These parameters are taken from Refs.~\cite{nucl_para1,nucl_para2,nucl_para3}. The number of minimum-bias events generated for each collision system is presented in Table~\ref{tab2}. 

\begin{table}[!htbp]
\begin{center}
\caption{Number of events (in millions) generated for various collision systems from the AMPT-SM model.}
\label{tab2}
\begin{adjustbox}{width=\columnwidth,center}
\begin{tabular}{ccccccc}
\hline
\hline
System 		& O+O	& Cu+Cu & Zr+Zr & Ru+Ru & Au+Au & U+U \\
\hline
Events (M)	& 50 	& 15 	& 9 	& 9 	& 6  	& 6	\\
\hline
\hline
\end{tabular}
\end{adjustbox}
\end{center}
\end{table}

\section{Analysis and results}
\label{subsec:analysis}
A comprehensive study has been conducted on the rapidity ($y$) dependence of $v_1$ for identified and charged hadrons at beam energies, ranging from $\sq$ = 7.7 to 200 GeV, utilizing the new coalescence AMPT model~\cite{newCoal,AMPTv1}. Another study performed a Coalescence Sum Rule (CSR) test involving seven produced hadrons using the same model~\cite{AMPTv1PLB,AMPT_v1_Univ}. In this paper, we report system size dependence of charged hadrons $v_1$ at $\sq$ = 200 GeV, using the same version of the new coalescence AMPT model.
\begin{figure*}[!htbp] 
\centering
\includegraphics[scale=0.55]{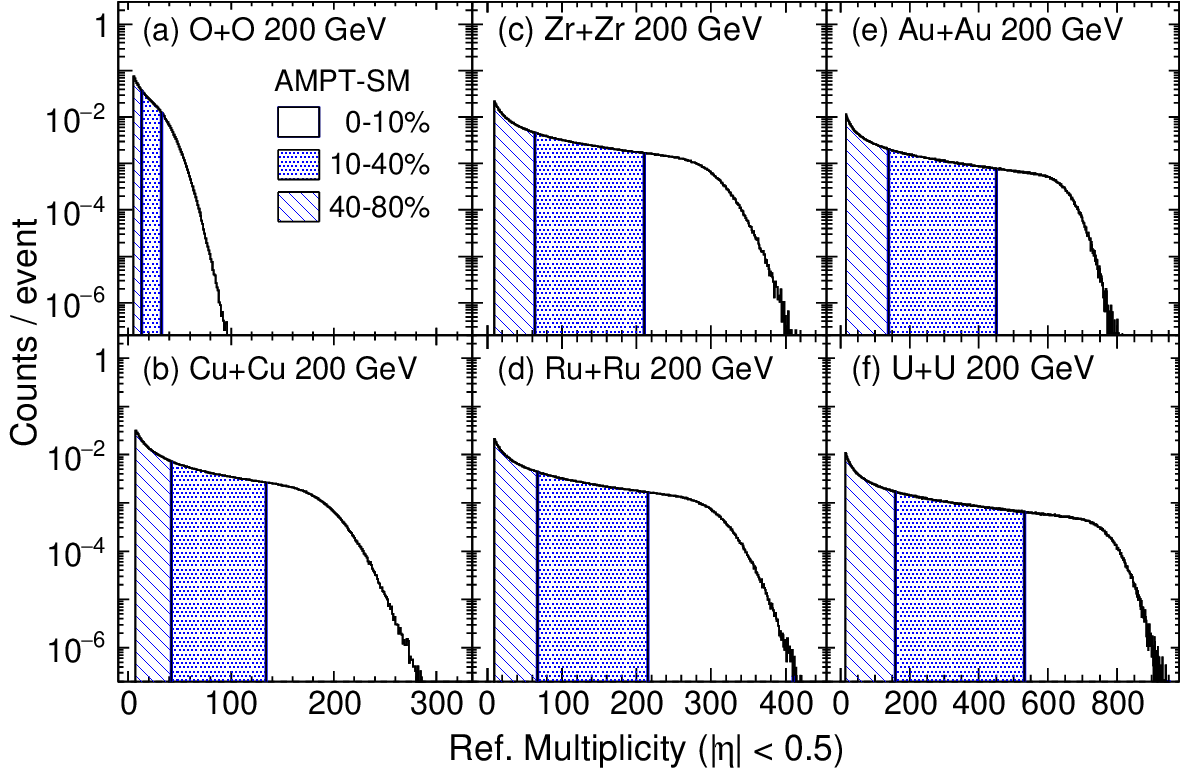}
\caption{(Color online) Charged particle multiplicity distributions within $|\eta| < 0.5$ for (a) O+O, (b) Cu+Cu, (c)  Ru+Ru, (d) Zr+Zr, (e) Au+Au, and (f) U+U collisions at $\sq$ = 200 GeV using the AMPT-SM model. The centrality classes 0-10\%, 10-40\%, and 40-80\% are shown in different bands.}
\label{fig:RefMult}
\end{figure*}

The analysis is performed across different centrality classes: 0-10\%, 10-40\%, and 40-80\%. This classification is based on the charged particle multiplicity ($\pi^{\pm}$, $K^{\pm}$, $p$, and $\bar{p}$) within $|\eta| < 0.5$, utilizing the AMPT-SM model similar to the experimental analysis. Figure~\ref{fig:RefMult} shows the multiplicity distribution for O+O, Cu+Cu, Ru+Ru, Zr+Zr, Au+Au, and U+U collisions at $\sq$ = 200 GeV. The different centrality classes are represented as bands. As expected, the highest multiplicity increases from approximately 100 to 900 with system size, enabling us to investigate the size dependence of $v_1$ across a broad range of multiplicity.

\subsection{$\eta$ dependence of $v_1$}
\label{ssec:v1eta}
The $v_1(\eta)$ may exhibit flatness at mid-rapidity due to the strong tilted expansion of the source, which can create an anti-flow perpendicular to the surface of the source~\cite{dflow4}. This anti-flow opposes the motion of nucleons, which could lead to a negative $\vslope$ if the tilted expansion is significant. Hydrodynamic models with a tilted source can describe the negative slope of $v_1$ observed at RHIC, which is especially notable at $\sq$ = 200 GeV~\cite{flow6}. Therefore, measuring $v_1(\eta)$ and $\vslope$ particularly in different $\pt$ ranges, which are related to soft and hard particle production processes, is crucial to understand the effect on the flow profile of the produced hadrons in different colliding systems.

\begin{figure*}[!htbp]
\centering
\includegraphics[scale=0.6]{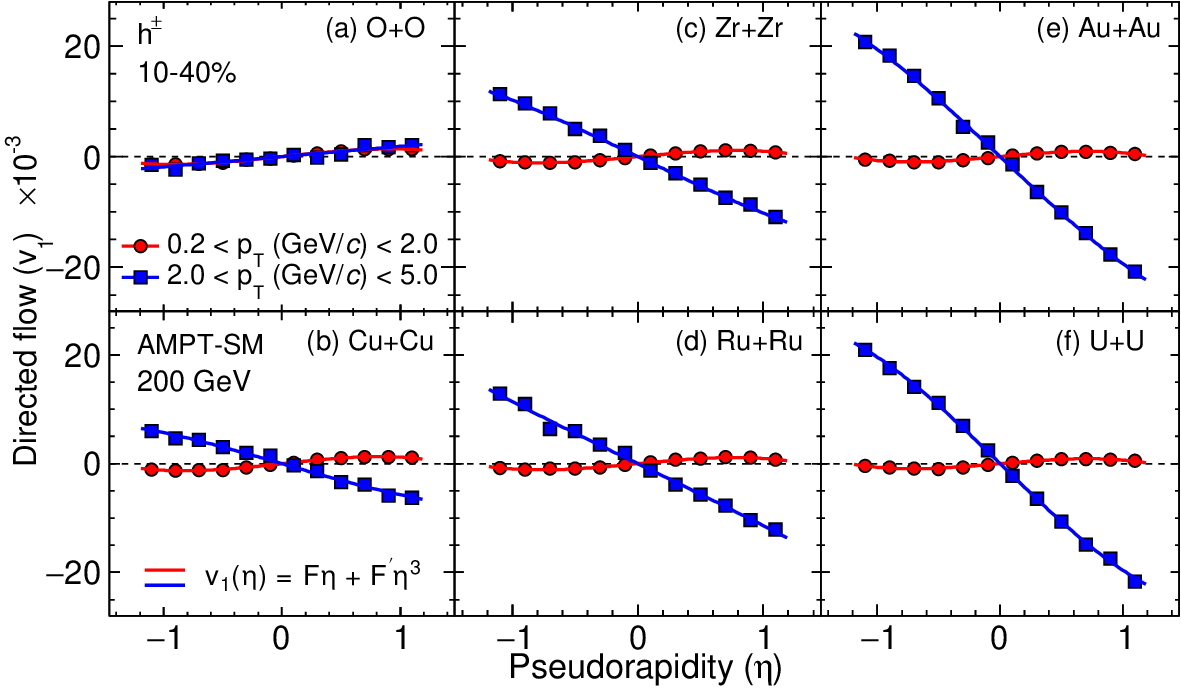}
\caption{$v_1(\eta)$ of charged hadrons in O+O, Cu+Cu, Ru+Ru, Zr+Zr, Au+Au, and U+U collisions for 10-40\% centrality at $\sq$ = 200 GeV using the AMPT-SM model. The red and blue markers correspond to low-$\pt$ (0.2 $< \pt <$ 2.0 GeV/$c$) and high-$\pt$ (2.0 $< \pt <$ 5.0 GeV/$c$), respectively. The solid lines indicate cubic polynomial function fit.}
\label{fig:v1vsEta}
\end{figure*}

We obtained charged hadron $v_1$ as a function of $\eta$ in 10-40\% central O+O, Cu+Cu, Ru+Ru, Zr+Zr, Au+Au, and U+U collisions at $\sq$ = 200 GeV using the AMPT-SM model. A comparison of $v_1(\eta)$ for low-$\pt$ (0.2 $< \pt <$ 2.0 GeV/$c$) and high-$\pt$ (2.0 $< \pt <$ 5.0 GeV/$c$) is  shown in Fig.~\ref{fig:v1vsEta}. The $v_1(\eta)$ is fitted with a cubic polynomial function of the form $v_1(\eta) = F\eta + F^{'}\eta^{3}$ within the fitting range of $|\eta| < 1.2$. This allows us to determine the slope parameter of $v_1$ at mid-rapidity, denoted as $F$ (= d$v_1$/d$\eta$). Our analysis shows a positive slope for $v_1$ among low-$\pt$ charged hadrons across all colliding systems. In contrast, the trend for $v_1$ at high-$\pt$ exhibits the opposite behavior compared to low-$\pt$ charged hadrons, with the exception for the smallest colliding system, O+O.    

In relativistic heavy-ion collisions, high-$\pt$ hadrons are less influenced by hadronic re-scattering due to their higher momentum. This enables them to escape the dense medium created during the collisions before significant interactions can occur~\cite{dflow8}. In contrast, low-$\pt$ hadrons experience more frequent interactions with the medium. As a result, while both high-$\pt$ and low-$\pt$ charged hadrons can be affected by the medium, the impact is less pronounced for high-$\pt$ hadrons. In Fig.~\ref{fig:v1vsEta}, we observe a larger magnitude of $v_1$ for high-$\pt$ charged hadrons across various systems, except in O+O collisions. This suggests that high-$\pt$ charged hadrons are better suited in retaining information about the initial dynamics compared to low-$\pt$ charged hadrons. The large difference in $v_1$ between high-$\pt$ and low-$\pt$ charged hadrons may also attributed to the tilt of the initial source.

\subsection{$v_1$-slope ($\vslope$)}
The nuclear passage time ($t_{pass}$) is affected by collision energy and system size. It decreases with a decrease in the mass or size of colliding nuclei. If nuclear passage time is comparable to the expansion time, spectator nucleons hinder the path of emitted hadrons toward the reaction plane. This shadowing effect caused by spectators can decrease $v_1$-slope with increasing mass of colliding nuclei at $\sq$ = 200 GeV~\cite{exd1}.

\begin{figure}[!htbp]
\centering
\includegraphics[scale=0.3]{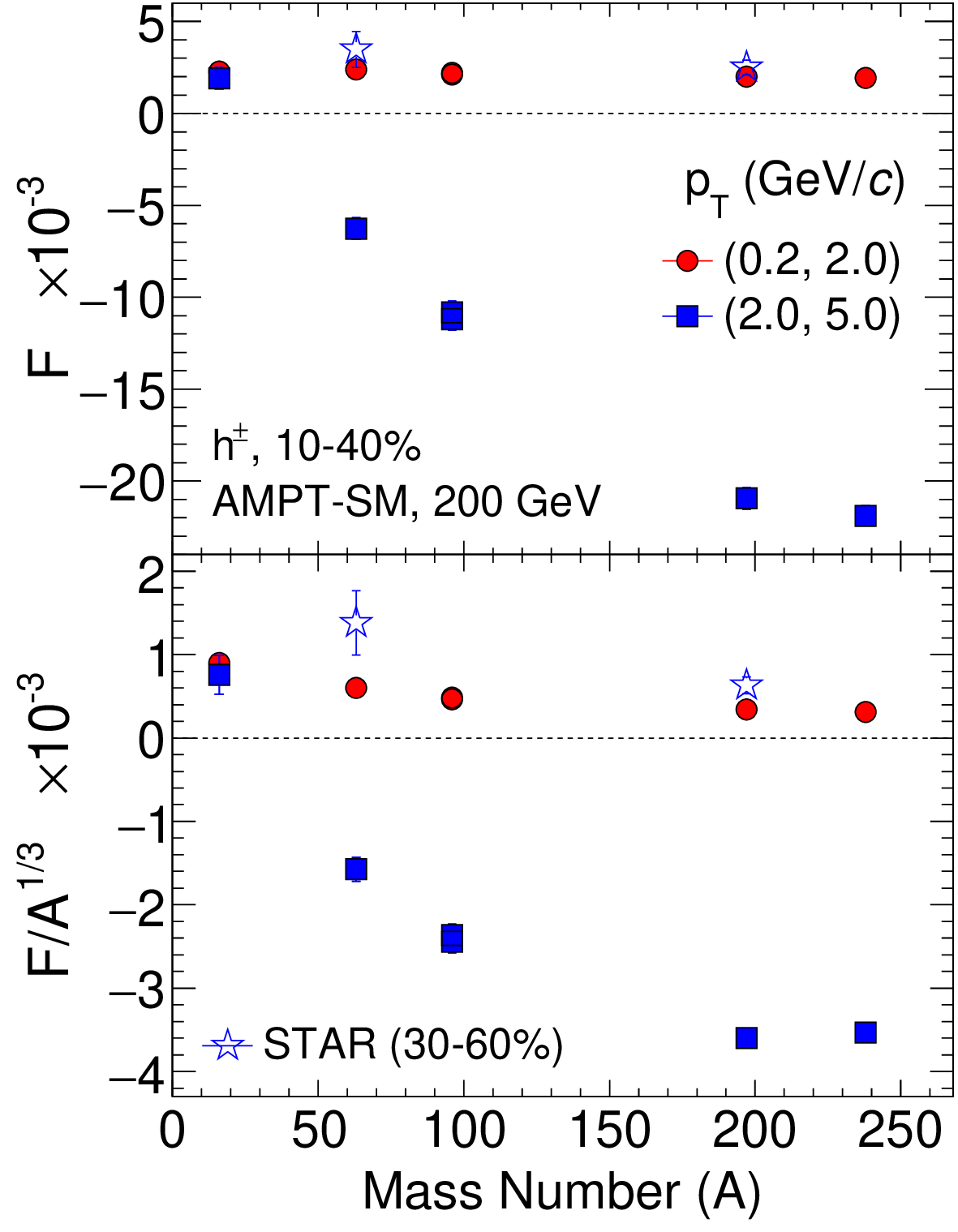}
\caption{The charged hadron $v_1$-slope parameter (F) and F/A$^{1/3}$ from the AMPT-SM model compared with the magnitude of the STAR measurements~\cite{exdflow3} for low-$\pt$ at $\sq$ = 200 GeV. The corresponding results for high-$\pt$ charged hadrons from the AMPT-SM model is also shown.}
\label{fig:v1slopeA}
\end{figure}

We present a comparison of charged hadrons $\vslope$ as a function of mass number ($A$) calculated from the AMPT-SM model with the magnitude of $\vslope$ from the STAR experiment at RHIC, as shown in Fig.~\ref{fig:v1slopeA}. The $\vslope$ for the experimental data is derived by fitting the published $v_1(\eta)$ results for the centrality 30-60\%, taken from the Ref.~\cite{exdflow3}. The fitting is performed using a cubic polynomial function in the same way as detailed in sub-section~\ref{ssec:v1eta}. Although the AMPT model results for centrality 10-40\% is compared, we have confirmed that the results from centrality 30-60\% also well describe the experimental data.
The STAR experiment reported a system size independence between Cu+Cu and Au+Au collisions, based on the measurements of charged hadron $v_1(\eta)$ at $\sq$ = 200 GeV~\cite{exdflow3}. The results from this study using the AMPT-SM model also support a similar conclusion of system size independence of $\vslope$ between the Cu+Cu and Au+Au collisions at $\sq$ = 200 GeV. However, it is important to note that comparing the $v_1$-slope among different systems rather than $v_1(\eta)$ provides a more effective way to assess the strength of $v_1$. 

In order to compare the flow results for different colliding systems, it was suggested to use the scaling factor $1.15(A_{P}^{1/3} + A_{T}^{1/3})$. Here, $A_{P}$ and $A_{T}$ correspond to the mass number of the projectile and target nuclei, respectively~\cite{exd2}. For symmetric colliding systems, where $A_{P} = A_{T} = A$, we present the slope parameter $F$ scaled by $A^{1/3}$, as a function of mass number. The scaled slope parameter $F/A^{1/3}$ reveals a clear system size dependence for both the low- and high-$\pt$ charged hadrons from the AMPT-SM model, as well as in the STAR experimental data. The relative percentage change of $F/A^{1/3}$ between Cu+Cu and Au+Au collisions for the experiment data is found to be 54.5$\pm$17.5 and the corresponding values from the AMPT-SM model is 42.3$\pm$1.9 for low-$\pt$ charged hadrons. The results for high-$\pt$ charged hadrons from the AMPT-SM model shows even larger relative percentage change of 128.5$\pm$12.5. 

\subsection{Centrality dependence of $v_1$-slope}
\begin{figure*}[!htbp]
\centering
\includegraphics[scale=0.6]{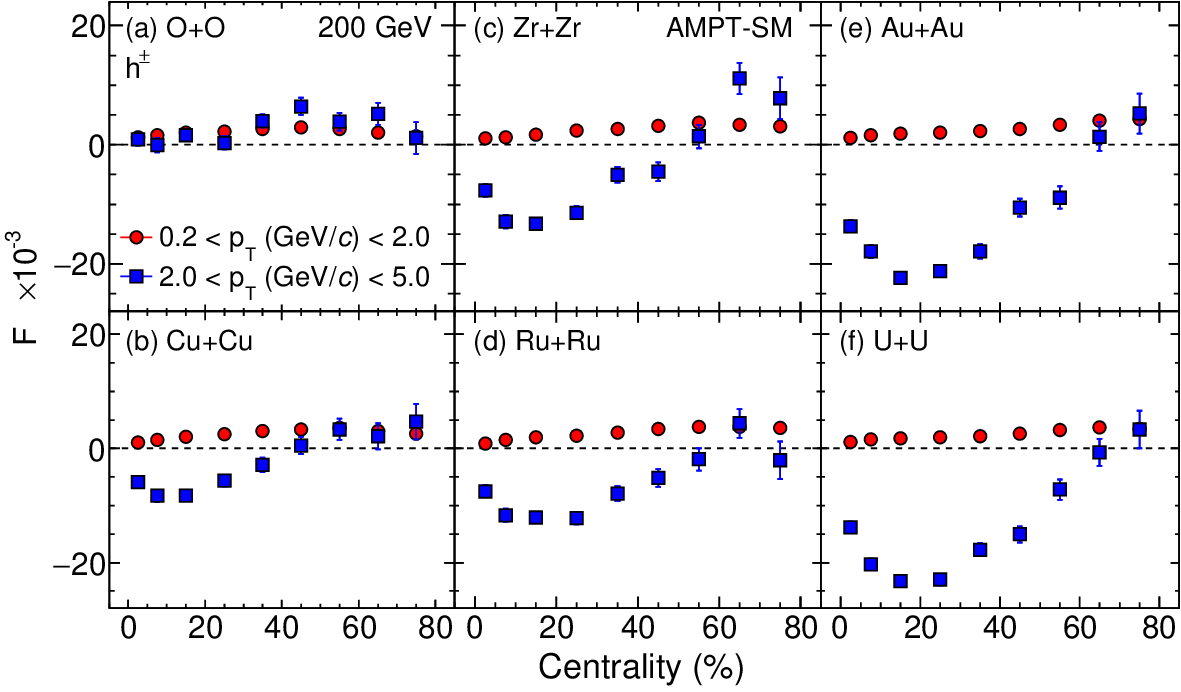}
\caption{The charged hadron $v_1$-slope as function of centrality (\%) in O+O, Cu+Cu, Ru+Ru, Zr+Zr, Au+Au and U+U collisions at $\sq$ = 200 GeV using the AMPT-SM model. The red and blue markers correspond to low-$\pt$ (0.2 $< \pt <$ 2.0 GeV/$c$) and high-$\pt$ (2.0 $< \pt <$ 5.0 GeV/$c$), respectively.}
\label{fig:v1slopeCent}
\end{figure*}
In symmetric nuclear collisions, centrality describes the degree of overlap between the colliding nuclei. It directly affects the initial geometry and the resulting collective flow. Generally, central collisions exhibit a weaker collective flow compared to peripheral collisions because the initial overlap area is more spatially symmetrical in central collisions. The impact of varying collision centrality has been suggested to provide control on the size of the participant fireball and the spectator fragment region~\cite{exd3}. Therefore, we study centrality dependence of $v_1$-slope for both low- and high-$\pt$ charged hadrons in different colliding systems using the AMPT-SM model. 

Figure~\ref{fig:v1slopeCent} shows the $v_1$-slope as a function of centrality in O+O, Cu+Cu, Ru+Ru, Zr+Zr, Au+Au, and U+U collisions at $\sq$ = 200 GeV. For low-$\pt$ charged hadrons, the $v_1$-slope increases from central to mid-central collisions and decreases towards peripheral collisions. In contrast, the $v_1$-slope for high-$\pt$ charged hadrons shows an opposite sign with larger magnitudes, except for the smallest system (O+O). The magnitude of $v_1$-slope initially increases from 0-20\% centrality, then decreases towards mid-central and peripheral collisions. The slope exhibits a crossover to positive values at mid-central or peripheral collisions, depending on the size of the colliding system. 

\subsection{System size dependence of $v_1$-slope}
The dependence of directed flow and its slope on the system size has been extensively studied at low beam energies ($\sim$0.1 to 2A GeV)~\cite{exd1,exd2,exd3,exd4,exd5}. At these lower energies, the slope of $v_1$ scaled with the radius parameter ($A_{P}^{1/3} + A_{T}^{1/3}$) has been discussed. The use of scaled variables simplifies the comparison of the $v_1$ among different colliding systems. 

\begin{figure*}[!htbp]
\centering
\includegraphics[scale=0.5]{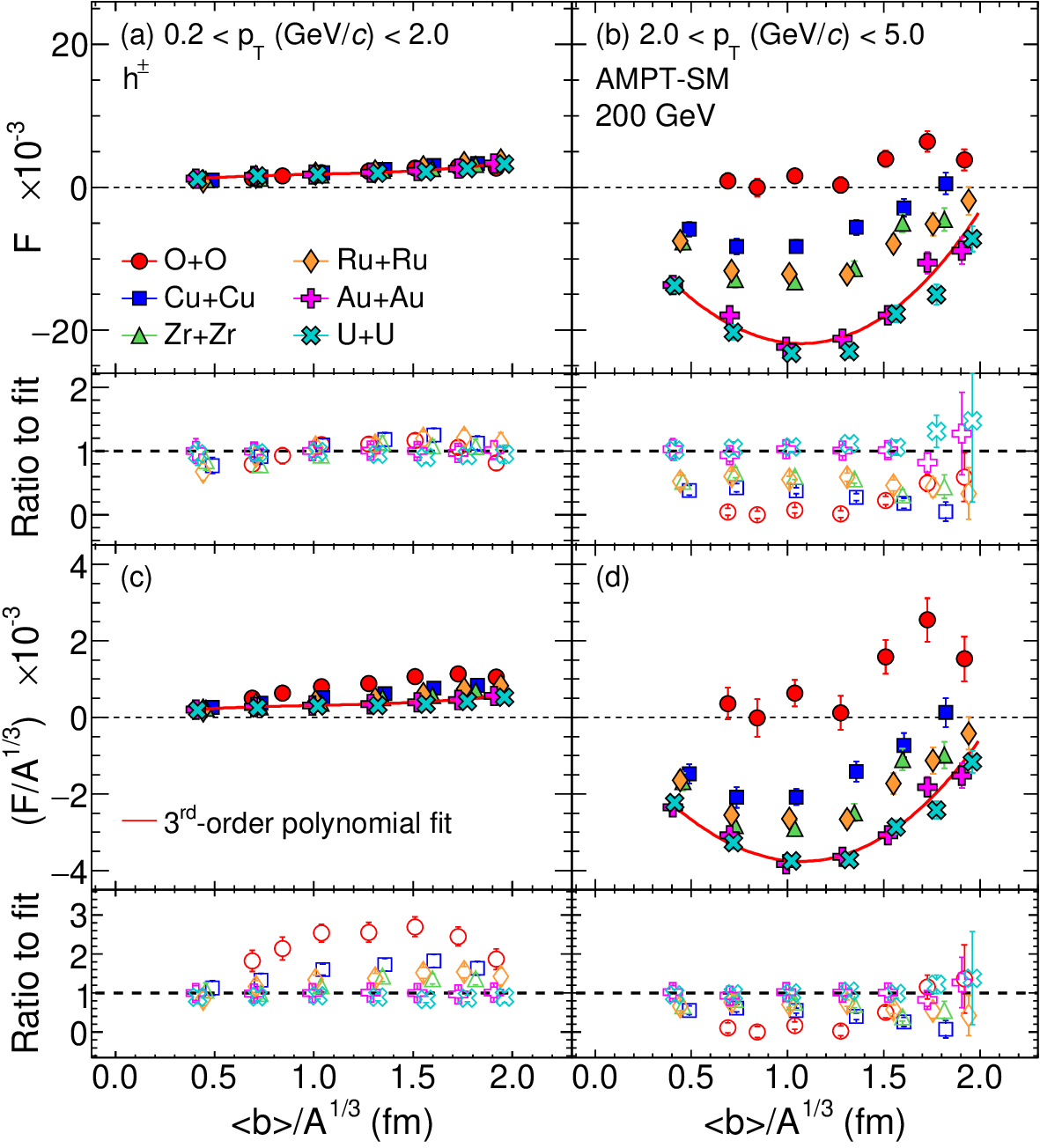}
\caption{The charged hadron $v_1$ slope ($F$) and scaled slope ($F/A^{1/3}$) parameter as function of $\langle b \rangle$/$A^{1/3}$ in O+O, Cu+Cu, Ru+Ru, Zr+Zr, Au+Au, and U+U collisions at $\sq$ = 200 GeV using the AMPT-SM model.}
\label{fig:Ratiov1slope}
\end{figure*}

Figure~\ref{fig:Ratiov1slope}(a,b) shows the $\vslope$ as function of scaled impact parameter ($\langle b \rangle/A^{1/3}$) for both the low-$\pt$ and high-$\pt$ charged hadrons in various colliding systems at $\sq$ = 200 GeV. The distribution for Au+Au collisions is fitted with a 3$^{rd}$-order polynomial function, and the corresponding ratios of $\vslope$ in other systems to the fit are shown in the lower panels of Fig.~\ref{fig:Ratiov1slope}(a,b). For low-$\pt$ charged hadrons, there is a consistent positive flow across all systems, indicating that they move along with the bulk matter. The ratios of $\vslope$ among different colliding systems indicate a negligible system size dependence within statistical uncertainties. In comparison, the $\vslope$ values for high-$\pt$ charged hadrons are negative and show a clear system size dependence. This indicates that high-$\pt$ charged hadrons flow in the opposite direction to the bulk matter. However, in the case of the smallest colliding system, O+O, there are no sufficient interactions to generate bulk matter to oppose the motion of high-$\pt$ hadrons. 

The scaling of the $v_1$-slope with system size has been proposed in Refs.~\cite{exd1,exd4}. Within an ideal hydrodynamics framework, the flow angle is expected to be a purely geometric quantity independent of the system size~\cite{exd6}. Any observed dependence on system size reflects the non-equilibrium nature of heavy-ion collisions. In Fig.~\ref{fig:Ratiov1slope}(c, d), the scaled $v_1$-slope ($F/A^{1/3}$) is presented as a function of $\langle b\rangle/A^{1/3}$ for various colliding systems at low-$\pt$ and high-$\pt$ using the AMPT-SM model. The scaling factor $A^{1/3}$ removes the contribution of system size. However, the $F/A^{1/3}$ does not show scaling, and the deviation increases as the size of the colliding system decreases for both low- and high-$\pt$ charged hadrons. 

\section{SUMMARY}
\label{sub:summary}
A comprehensive study has been conducted on directed flow ($v_1$) and its slope at mid-rapidity ($\vslope$) for low- and high-$\pt$ charged hadrons ($h^{\pm}$) in various colliding systems (O+O, Cu+Cu, Ru+Ru, Zr+Zr, Au+Au, and U+U) at $\sq =$ 200 GeV using the new coalescence AMPT-SM model. The observed positive $v_1$-slope ($\vslope > 0$) for low-$\pt$ charged hadrons suggest that soft particles move along with the bulk medium. However, high-$\pt$ charged hadrons flow in the opposite direction to the bulk medium, resulting in a negative $v_1$-slope ($\vslope < 0$)  for all colliding systems except for O+O collisions. In the smallest colliding system (O+O), it appears that the QGP bulk may not generate sufficient interactions to oppose the motion of high-$\pt$ charged hadrons. In larger colliding systems, the opposite sign of $v_1$-slope for low- and high-$\pt$ charged hadrons suggests a hard and soft asymmetry in the particle production profile.

A centrality dependence of directed flow slope is observed for low- and high-$\pt$ charged hadrons across various colliding systems at $\sq$ = 200 GeV. A positive slope is found in both small systems, such as O+O, across all centralities, and in peripheral collisions of heavier systems. This trend suggests an increasing contribution from spectators in generating the positive directed flow of charged hadrons. Furthermore, the centrality dependence of the $v_1$-slope could also indicate an increasing contribution of transported quarks from central to peripheral collisions.

A system size independence of $\vslope$ at low-$\pt$ is found for charged hadrons between Au+Au and Cu+Cu collisions. This finding is consistent with measurements from the STAR experiment at RHIC. However, a better way of comparing directed flow is via its slope $\vslope$ rather than $v_1(\eta)$, and to see the effect of system size, one should compare the $v_1$-slope scaled by the system size, ($\vslope$)/$A^{1/3}$. At high-$\pt$, the charged hadron directed flow and its slope depend on the size of the colliding system, which indicates a violation of the entropy-driven multiplicity scaling. The predictions reported here, based on the AMPT-SM model regarding system size dependence, can be tested in experiments with high statistics and precision data sets that can be used to constrain various model parameters. 
 
\section*{Acknowledgments}
\label{acknowledgement}
KN is supported by OSHEC, Department of Higher Education, Government of Odisha, Index No. 23EM/PH/124 under MRIP 2023. The authors thank Prof. B. Mohanty for providing computational facilities at NISER, India. The authors also thank Prof. Z.-W. Lin for the new coalescence AMPT model.

\end{document}